\documentclass[twocolumn]{aastex631}

\usepackage{graphicx}
\usepackage{amsmath}
\usepackage{natbib}
\usepackage{amssymb}

\newcommand{\pfrac}[2]{\left( \frac{#1}{#2} \right)}
\newcommand{\be}{\begin{equation}}
\newcommand{\ee}{\end{equation}}

\defcitealias{Linial&Metzger23}{LM23}

\submitjournal{ApJL}

\shorttitle{Ultraviolet Quasi-Periodic Eruptions}
\shortauthors{Linial \& Metzger}

\begin{document}

\title{Ultraviolet Quasi-periodic Eruptions from Star-Disk Collisions in Galactic Nuclei}

\correspondingauthor{Itai Linial}
\email{itailin@ias.edu, il2432@columbia.edu}

\author[0000-0002-8304-1988]{Itai Linial}
\affil{Institute for Advanced Study, 1 Einstein Drive, Princeton, NJ 08540, USA}
\affil{Department of Physics and Columbia Astrophysics Laboratory, Columbia University, New York, NY 10027, USA}

\author[0000-0002-4670-7509]{Brian D.~Metzger}
\affil{Department of Physics and Columbia Astrophysics Laboratory, Columbia University, New York, NY 10027, USA}
\affil{Center for Computational Astrophysics, Flatiron Institute, 162 5th Ave, New York, NY 10010, USA} 

\begin{abstract}
    ``Quasi-periodic eruptions'' (QPE) are recurrent nuclear transients with periods of several hours to almost a day, which thus far have been detected exclusively in the X-ray band. We have shown that many of the key properties of QPE flares (period, luminosity, duration, emission temperature, alternating long-short recurrence time behavior, source rates) are naturally reproduced by a scenario involving twice-per-orbit collisions between a solar-type star on a mildly eccentric orbit, likely brought into the nucleus as an extreme mass-ratio inspiral (EMRI), and the gaseous accretion disk of a supermassive black hole (SMBH).  The flare is generated by the hot shocked debris expanding outwards from either side of the disk midplane, akin to dual miniature supernovae.  Here, we consider the conditions necessary for disk-star collisions to generate lower-temperature flares which peak in the ultraviolet (UV) instead of the X-ray band.  We identify a region of parameter space at low SMBH mass $M_{\bullet} \sim 10^{5.5}M_{\odot}$ and QPE periods $P \gtrsim 10$ hr for which the predicted flares are sufficiently luminous $L_{\rm UV} \sim 10^{41}$ erg s$^{-1}$ to outshine the quiescent disk emission at these wavelengths.  The prospects to discover such ``UV QPEs'' with future satellite missions such as ULTRASAT and UVEX depends on the prevalence of very low-mass SMBH and the occurrence rate of stellar EMRIs onto them.  For gaseous disks produced by the tidal disruption of stars, we predict that X-ray QPEs will eventually shut off, only to later reappear as UV-QPEs as the accretion rate continues to drop.
\end{abstract}

\keywords{Supermassive black holes (1663), Tidal disruption (1696), Ultraviolet transient sources (1854), X-ray transient sources (1852)}

\section{Introduction}

Over recent years, a growing number of periodically flaring sources have been observed from the nuclei of distant galaxies, with recurrence times ranging from a few to tens of hours (Quasi-periodic Eruptions, or QPEs; e.g., \citealt{Miniutti+19,Giustini+20,Arcodia+21,Chakraborty+21}), to  weeks (e.g., \citealt{Guolo+23}), up to years or longer (e.g., \citealt{Payne+21,Wevers+22,Liu+22,Malyali+23}). While the physical explanation(s) for these ``periodic nuclear transients'' remains under debate, deciphering their mysteries offers the potential to unlock exciting new probes of the dynamics of stars and compact objects in close proximity to the supermassive black hole (SMBH) and its innermost accretion flow.

The short ($\lesssim $ several hours-long) flares from X-ray QPE systems are characterized by peak luminosities $\gtrsim 10^{42}$ erg s$^{-1}$ in the 0.5-2 keV X-ray band, at least an order of magnitude higher than the quiescent X-ray luminosity.  The flare spectra are quasi-thermal, with temperatures $\approx 100-200$ eV \citep{Miniutti+19,Giustini+20,Arcodia+21,Chakraborty+21,Arcodia+22,Miniutti+23,Webbe&Young23}. The low stellar masses of their host galaxies point to SMBHs of relatively low masses, $M_{\bullet} \lesssim 10^{6.5}M_{\odot}$ (e.g., \citealt{Wevers+22}).   

Many of the models proposed for X-ray QPEs involve the partial disruption or interaction with an accretion flow of a star or compact object in orbit around the SMBH \citep{Zalamea+10,King20,Sukova+21,Metzger+22,Zhao+22,King22,Krolik&Linial22,LuQuataert23,Linial&Sari17,Linial&Metzger23,Franchini+23,Tagawa&Haiman23}. However, only a few of these models can explain an important clue: a regular alternating behavior, observed in at least two of the QPE sources$-$GSN 069 \citep{Miniutti+19,Miniutti+23} and eRO-QPE2 \citep{Arcodia+21}, in which the temporal spacing between consecutive bursts varies back and forth by around 10\%.  Flares that precede longer recurrence intervals also appear systematically brighter than those appearing before short ones (e.g., \citealt{Miniutti+23}).

Recently, \citet[hereafter \citetalias{Linial&Metzger23}]{Linial&Metzger23} showed that many if not all of the observed properties of X-ray QPES (period, luminosity, duration, emission temperature, occurrence rates in galactic nuclei) can be reproduced in a scenario in which a main-sequence star on a mildly eccentric inclined orbit (an extreme mass-ratio inspiral; EMRI) collides twice per orbit with a gaseous accretion disk (see also \citealt{Xian+21,Tagawa&Haiman23,Franchini+23}). Such mildly eccentric stellar EMRIs are expected to be relatively common in galactic nuclei on orbital periods of several hours (e.g., \citealt{Linial&Sari22}), while the gaseous accretion disk is either produced by mass stripped by the star itself during this process (e.g., \citealt{LuQuataert23}) or was created by a recent but otherwise unrelated tidal disruption event (TDE) involving a different star (e.g., \citealt{Miniutti+19}). The latter may not be coincidental: the average interval between TDEs in a galactic nucleus is less than the EMRI's gravitational inspiral time across the expected radial scale of the TDE accretion disk (\citetalias{Linial&Metzger23}).  The oscillating long-short recurrence time behavior follows naturally in this scenario, due to the longer time the star spends between collisions on the apocenter side of the disk. 

In the \citetalias{Linial&Metzger23} scenario, the flares are powered by hot shocked disk material which expands from either side of midplane, akin to dual miniature supernova explosions (e.g., \citealt{Ivanov+98}).  The flare duration is set by the photon diffusion time through the expanding debris cloud, while the radiated energy is set by the thermal energy deposited by the star-driven shock, accounting for adiabatic losses from the collision site to larger radii in the outflow where radiation is no longer trapped. As in the case of supernova shock break-out, gas and radiation in the expanding debris may not be in equilibrium. In particular, because of its low density and rapid expansion rate, inefficient photon production in the debris can result in harder temperatures than the blackbody value (e.g., \citealt{Nakar&Sari10}). Such high emission temperatures are important to the detectability of X-ray QPEs, as they enable the harder flare emission to stick out above the softer quiescent disk emission (e.g., \citealt{Miniutti+23}). However, for different values of the SMBH mass, or properties of the accretion disk or star, photon production can be more efficient, resulting in the escaping radiation being considerably softer. In particular, there is no reason {\it a priori} within the \citetalias{Linial&Metzger23} model why QPE flares could not occur in the ultraviolet (UV) rather than X-ray band, where the quiescent accretion disk emission is comparatively dimmer.

A number of wide-field UV time-domain surveys are planned over the next decade.  Most notably, the Ultraviolet Transient Astronomy Satellite (ULTRASAT; \citealt{Sagiv+14}), will provide high etendue monitoring of the UV sky following its planned launch in 2025.  The Czech UV satellite mission QUVIK \citep{Werner+23} will provide similar capabilities, though with a shorter wavelength sensitivity and somewhat smaller field of view. The proposed space mission Ultraviolet Explorer (UVEX; \citealt{Kulkarni+21}) will perform a cadenced all-sky survey with greater sensitivity and covering a wider UV wavelength range.  These efforts make it an ideal time to explore the conditions under which ``UV-QPEs'' are predicted by the star-disk collision scenario.

This paper is organized as follows.  In Sec.~\ref{sec:model} we summarize the key results of \citetalias{Linial&Metzger23} for the properties of disk-star collision flares.  In Sec.~\ref{sec:detection} we consider the conditions to observe UV-QPE flares and consider their prospects for detection with future UV satellite missions. In Sec.~\ref{sec:discussion} we discuss our results and conclude.

\section{Disk-Star Collision Flare Properties }
\label{sec:model}

We consider a supermassive black hole (SMBH) of mass $M_{\bullet} = 10^{6}M_{\bullet,6}M_{\odot}$ accreting gas steadily at a rate $\dot{M} = \dot{m}\dot{M}_{\rm Edd}$, where
$\dot{M}_{\rm Edd} \equiv L_{\rm edd}/(\epsilon c^{2})$ is the Eddington accretion rate for a characteristic radiative efficiency of $\epsilon = 0.1$ and $L_{\rm Edd} \simeq 1.5\times 10^{44}M_{\bullet,6}$ erg s$^{-1}$.  In the inner regions of the disk of greatest interest, radiation pressure dominates over gas pressure and vertical aspect ratio at radii $r \gg R_{\rm g} \equiv GM_{\bullet}/c^{2}$ can be written (e.g., \citealt{Frank+02}) 
\be
\frac{h}{r} \simeq \frac{3}{2\epsilon}\frac{R_{\rm g}}{r}\frac{\dot{M}}{\dot{M}_{\rm Edd}} \simeq 1.5\times 10^{-2} \; \dot{m}_{-1}\left(\frac{r}{100R_{\rm g}}\right)^{-1},
\label{eq:hoverr}
\ee
where $h$ is the vertical scale-height and $\dot{m} = 0.1\dot{m}_{-1}$.  The optical depth through the disk midplane of surface density $\Sigma \simeq \dot{M}/(3\pi \nu)$ can be written
\be
\tau_{\rm c} = \Sigma \kappa_{\rm T} \simeq \frac{\dot{M}\kappa_{\rm T}}{3\pi \nu} \approx \frac{6.0\times 10^{3}}{\alpha_{-1}\dot{m}_{-1}}\left(\frac{r}{100R_{\rm g}}\right)^{3/2},
\label{eq:tauc}
\ee
where $\kappa_{\rm T} \simeq 0.34$ cm$^{2}$ g$^{-1}$ is the electron scattering opacity, $\nu = \alpha (GM_{\bullet} r)^{1/2}(h/r)^{2}$ is the kinematic viscosity \citep{Shakura&Sunyaev73}, and we scale $\alpha = 0.1\alpha_{-1}$ to a characteristic value.  Here we have assumed the disk midplane is radiation dominated, which is valid at radii obeying,
\be
\frac{r}{R_{\rm g}} \lesssim 450 \; ( M_{\bullet,6} \alpha_{-1} )^{2/21} \dot{m}_{-1}^{16/21} \,,
\ee
corresponding to QPE periods $\lesssim$ 1 day for typical parameters (see below).

Absent interaction with orbiting bodies (i.e., when in ``quiescence''), the disk emission is dominated by radii near the innermost circular orbit $R_{\rm isco}$, with total luminosity
\be
L_{\rm Q} = \dot{m}L_{\rm Edd} \simeq 1.5\times 10^{43}\,{\rm erg\,s^{-1}}\dot{m}_{-1}M_{\bullet,6}
\label{eq:LQ},
\ee
and characteristic emission temperature
\be
k_{\rm B}T_{\rm Q} \approx k_{\rm B}\left(\frac{3GM_{\bullet}\dot{M}}{8\pi \sigma R_{\rm isco}^{3}}\right)^{1/4} \simeq 59\,{\rm eV}\,\frac{\dot{m}_{-1}^{1/4}}{M_{\bullet,6}^{1/4}}\left(\frac{R_{\rm isco}}{4R_{\rm g}}\right)^{-3/4}.
\label{eq:TQ}
\ee
The origin of the quiescent disk, namely whether it is fed by gas stripped from the EMRI itself near the collision radius (e.g., \citealt{LuQuataert23}), from an unrelated TDE (e.g., \citetalias{Linial&Metzger23}), or as a part of a more radially-extended accretion flow into the galactic nucleus, is uncertain.  Nevertheless, on the modest radial scales which produce most of the disk's UV emission, a steady accretion flow onto the SMBH is a reasonable approximation.  Assuming multi-color blackbody thin-disk emission, the luminosity at a characteristic UV frequency $h \nu \ll k_{\rm B}T_{\rm Q}$ can approximately be written\footnote{The dimensionless prefactor in equation \ref{eq:LQnu} is given by $\frac{120}{\pi^4} \, \left( \int_0^\infty x^{5/3}/(e^x-1) \, dx \right) \approx 2.38$.}
\begin{multline}
    \nu L_{\rm Q,\nu} \simeq 2.38 \cos i \, L_{\rm Q}\left(\frac{h\nu}{k_{\rm B}T_{\rm Q}}\right)^{4/3} \approx \\
    1.5\times 10^{42}\,{\rm erg\,s^{-1}}\; \cos{i} \, \dot{m}_{-1}^{2/3}M_{\bullet,6}^{4/3}\left(\frac{h\nu}{5\,{\rm eV}}\right)^{4/3},
\label{eq:LQnu}
\end{multline}

where $i$ is the viewer inclination relative to the angular momentum axis of the disk and in the second equality we have assumed $R_{\rm isco} = 4 R_{\rm g}$. If the disk is fed by gas at the stellar collision point, such that the disk spreads radially outwards from that point, its spectrum can be somewhat steeper than shown in Eq.~\eqref{eq:LQnu}, $\nu L_{\rm Q,\nu} \propto \nu^{7/3}$ \citep{LuQuataert23}.  However, for typical parameters the radii of the quiescent disk which contribute most of the UV emission reside only moderately outside the EMRI collision radius\footnote{
The quiescent disk emission at reference frequency $\nu$ is dominated by an annulus of radius \begin{equation}
    r(\nu) \approx \pfrac{3G M_\bullet \dot{M}}{8 \pi \sigma (h\nu/(A_{\rm BB} k_{\rm B}))^4}^{1/3} \,,
\end{equation}
where $A_{\rm BB} \approx 3$, corresponding to the peak of the blackbody emission, $h\nu_{\rm max} \approx 3 k_{\rm B} T$. Relative to the star collision radius
\begin{multline}
    \frac{r(\nu)}{r_0} \approx \pfrac{45 A_{\rm BB}^4}{16\pi^4}^{1/3} \pfrac{\dot{M} c^2}{ h \nu^4 P_{\rm QPE}^{2}}^{1/3} \approx \\
    5 \; 
    \pfrac{M_{\bullet,6} \dot{m}_{-1}}{\mathcal{P}_{\rm QPE,4}^2 }^{1/3} \pfrac{h\nu}{5 \, \rm eV}^{-4/3} \,.
\end{multline}}, resulting in only a modest correction to the UV luminosity of the disk relative to that predicted by Eq.~\eqref{eq:LQnu}.

The colliding body is fiducially taken to be a star of radius $R_{\star} = \mathcal{R}_{\star} R_{\odot}$ and mass $M_{\star} = \mathcal{M}_{\star} M_{\odot}$ on a mildly eccentric orbit around the SMBH of semi-major axis $r_{0}$.  The period between flares, $P_{\rm QPE}$, is set by collisions between the star and the disk which occur twice per orbital period $P_{\rm orb} = 2\pi(r_0^{3}/GM_{\bullet})^{1/2}$,  
\be
P_{\rm QPE} \approx \frac{P_{\rm orb}}{2} \simeq 4.3\,{\rm hr}\,M_{\bullet,6}\left(\frac{r_{0}}{100R_{\rm g}}\right)^{3/2}.
\label{eq:TQPE}
\ee

The condition that the collision radius must exceed the star's Roche radius, $r_0 \ge r_{\rm T} \simeq 2R_{\star}(M_{\bullet}/M_{\star})^{1/3}$, defines a minimum QPE period
\be
P_{\rm QPE,min} \simeq \pi \left(\frac{8R_{\star}^{3}}{GM_{\star}}\right)^{1/2} \simeq 3.9\,{\rm hr}\,\mathcal{R}_{\star}^{3/2}\mathcal{M}_{\star}^{-1/2}\,.
\label{eq:TQPEmin}
\ee
We expect $P_{\rm QPE} \gtrsim P_{\rm QPE,min}$ if the gaseous accretion disk which enables the flares arises from the tidal disruption of a star similar to the EMRI (\citetalias{Linial&Metzger23}), for which the circularization radius of the stellar debris is comparable to the tidal radius.  

The star will twice per orbit pass through the disk.  For the most common case of a roughly head-on collision, the collision speed is roughly equal to the Keplerian orbital velocity $v_{\rm c} \sim v_{\rm K} \approx (GM_{\bullet}/r_0)^{1/2}$ and the mass of the disk material intercepted each passage is $M_{\rm ej} \simeq 2\pi R_{\star}^{2}\Sigma$.  The shocked ejecta is initially highly optically thick (optical depth $\tau_{\rm c} \gg 1$; Eq.~\eqref{eq:tauc}) and will emerge from each side of the disk spreads roughly spherically in all directions (e.g., \citealt{Ivanov+98}), achieving an asymptotic velocity $v_{\rm c} \sim v_{\rm ej} \sim v_{\rm K}$.  The timescale of the resulting flare is given by when the optical depth of the ejecta $\tau \propto t^{-2}$ achieves $\tau \sim c/v_{\rm ej}$, as occurs on a timescale
\be
t_{\rm QPE} \approx \left(\frac{\kappa M_{\rm ej}}{4\pi c v_{\rm ej}}\right)^{1/2} \approx 
0.09\,{\rm hr}\frac{\mathcal{R}_{\star}\mathcal{P}_{\rm QPE,4}^{2/3}}{\alpha_{-1}^{1/2}\dot{m}_{-1}^{1/2}M_{\bullet,6}^{2/3}},  \,
\label{eq:tpk}
\ee
and corresponding radius $R_{\rm diff} \simeq v_{\rm K}t_{\rm QPE}.$  Within a factor of a few, Eq.~\eqref{eq:tpk} sets the rise time and decay time, as well as the overall duration, of the bolometric light curve. 
The flare luminosity is set by the thermal energy which remains in the shocked ejecta after expanding adiabatically from the midplane to the diffusion surface (where $\tau \sim c/v_{\rm ej}$) and can be written (\citetalias{Linial&Metzger23})
\begin{multline}
    L_{\rm QPE} \approx \frac{L_{\rm Edd}}{3}\frac{(R_\star^2 h)^{1/3} }{r_{0}} 
    \approx \\
    3.5 \times 10^{41}\,{\rm erg\,s^{-1}}\,\mathcal{R}_{\star}^{2/3}M_{\bullet,6}\dot{m}_{-1}^{1/3}\mathcal{P}_{\rm QPE,4}^{-2/3}.
    \label{eq:Lpk}
\end{multline}
The blackbody temperature of the radiation is set by the energy density of the radiation at the diffusion surface and can be written
\be
k_{\rm B} T_{\rm BB} \approx 12.6\,{\rm eV} \frac{\alpha_{-1}^{1/4}\dot{m}_{-1}^{1/3}M_{\bullet,6}^{1/3}}{\mathcal{R}_{\star}^{1/3}\mathcal{P}_{\rm QPE,4}^{1/4}}   \,.
\label{eq:TBB}
\ee
In general the temperature of the escaping photons $T_{\rm obs}$ can be harder than $T_{\rm BB}$ if photon production via free-free emission in the expanding material is not rapid enough to achieve thermal equilibrium in the ejecta on the timescale of the emission (e.g., \citealt{Weaver76,Katz+10,Nakar&Sari10}).  This occurs when (\citetalias{Linial&Metzger23})
\be \label{eq:eta}
\eta \approx
2.46 \alpha_{-1}^{9/8}\dot{m}_{-1}^{5/4}M_{\bullet,6}^{13/6}\mathcal{P}_{\rm QPE,4}^{-49/24} \gg 1.
\ee
In general, we have
\be
 k_{\rm B}T_{\rm obs} \approx \,{\rm max}[1,\eta^{2}]k_{\rm B}T_{\rm BB} \,,
\label{eq:Tobs}
\ee
and explictly, for $\eta>1$\footnote{Eq.~\ref{eq:Tobs_eta>1} appears in \citetalias{Linial&Metzger23} as their Eq. 21, however, with a typo in the scaling of $T_{\rm obs}$ with $\dot{m}$ ($\dot{m} \propto \dot{m}^{11/4}$ in \citetalias{Linial&Metzger23} instead of $\dot{m}^{17/6}$). The difference in scaling (of $\dot{m}^{1/12}$) is weak and does not affect any of the conclusions.}
\begin{equation} \label{eq:Tobs_eta>1}
    \left. k_{\rm B}T_{\rm obs} \right|_{\eta>1}
    \approx 76 \, {\rm eV} \; \frac{\alpha_{-1}^{5/2} \dot{m}_{-1}^{17/6} M_{\bullet,6}^{14/3}}{\mathcal{R}_\star^{1/3} P_{\rm QPE,4}^{13/3}} \,.
\end{equation}
In summary, the disk-collision model predicts the observables $\{t_{\rm QPE}, L_{\rm QPE}, T_{\rm obs}\}$ in terms of $\{\mathcal{R}_{\star},\dot{m},\alpha,\mathcal{P}_{\rm QPE} \gtrsim \mathcal{P}_{\rm QPE,min}(M_{\star})\}$. 

\begin{figure*}
    \includegraphics[width=\textwidth]{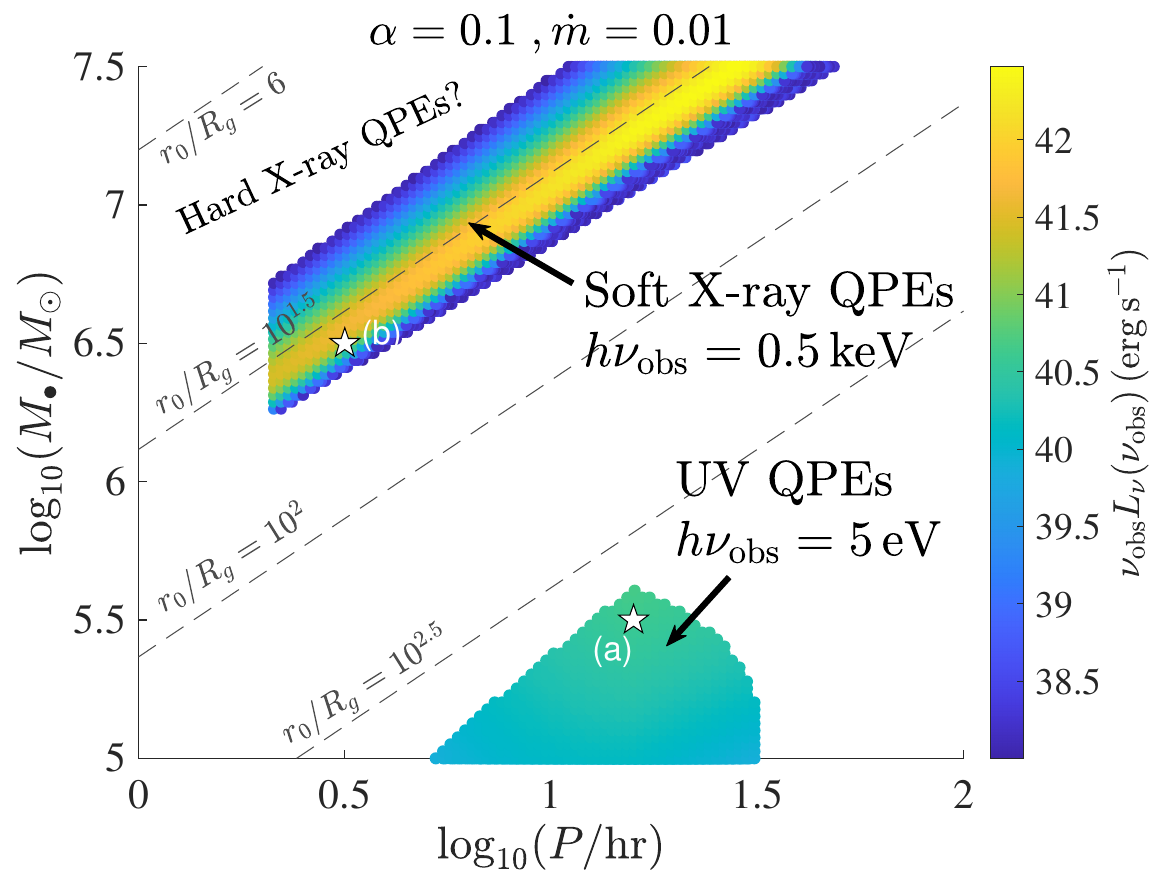}        \includegraphics[width=0.49\textwidth]{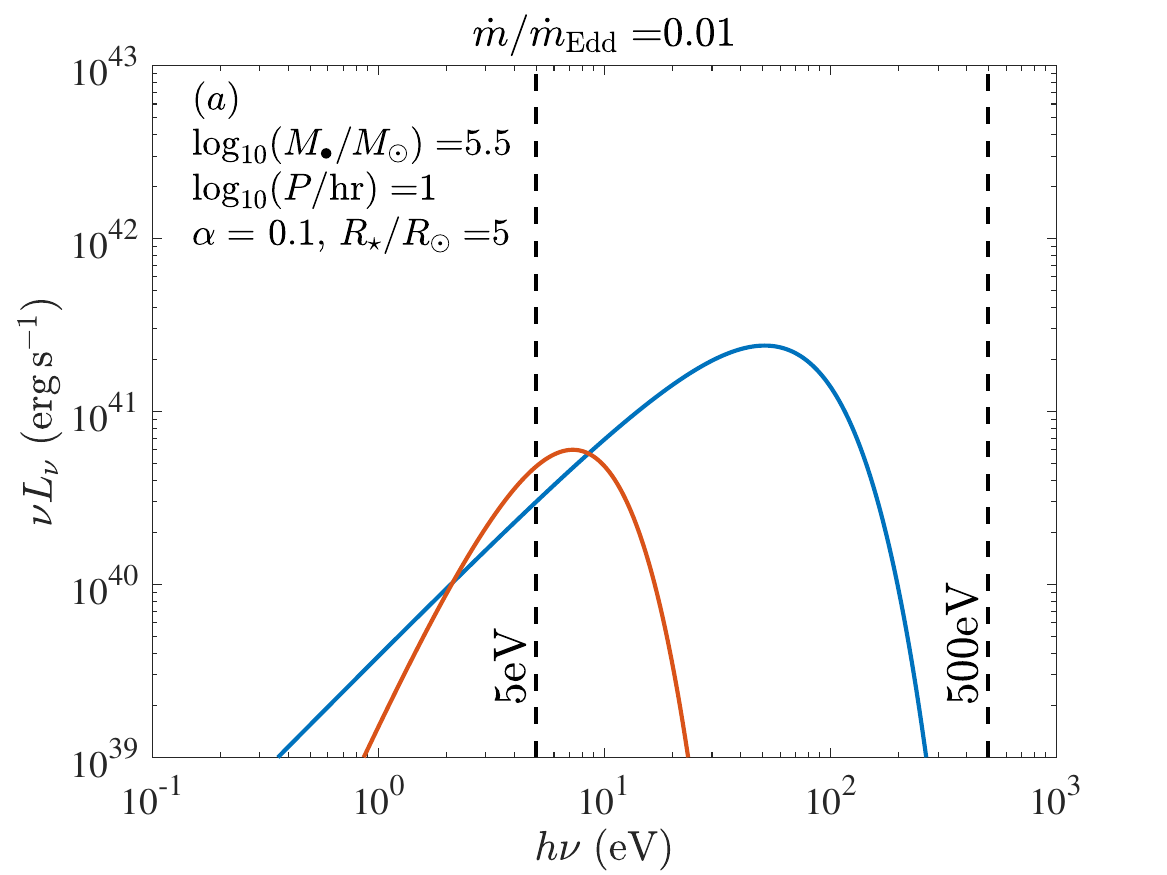}
    \includegraphics[width=0.49\textwidth]{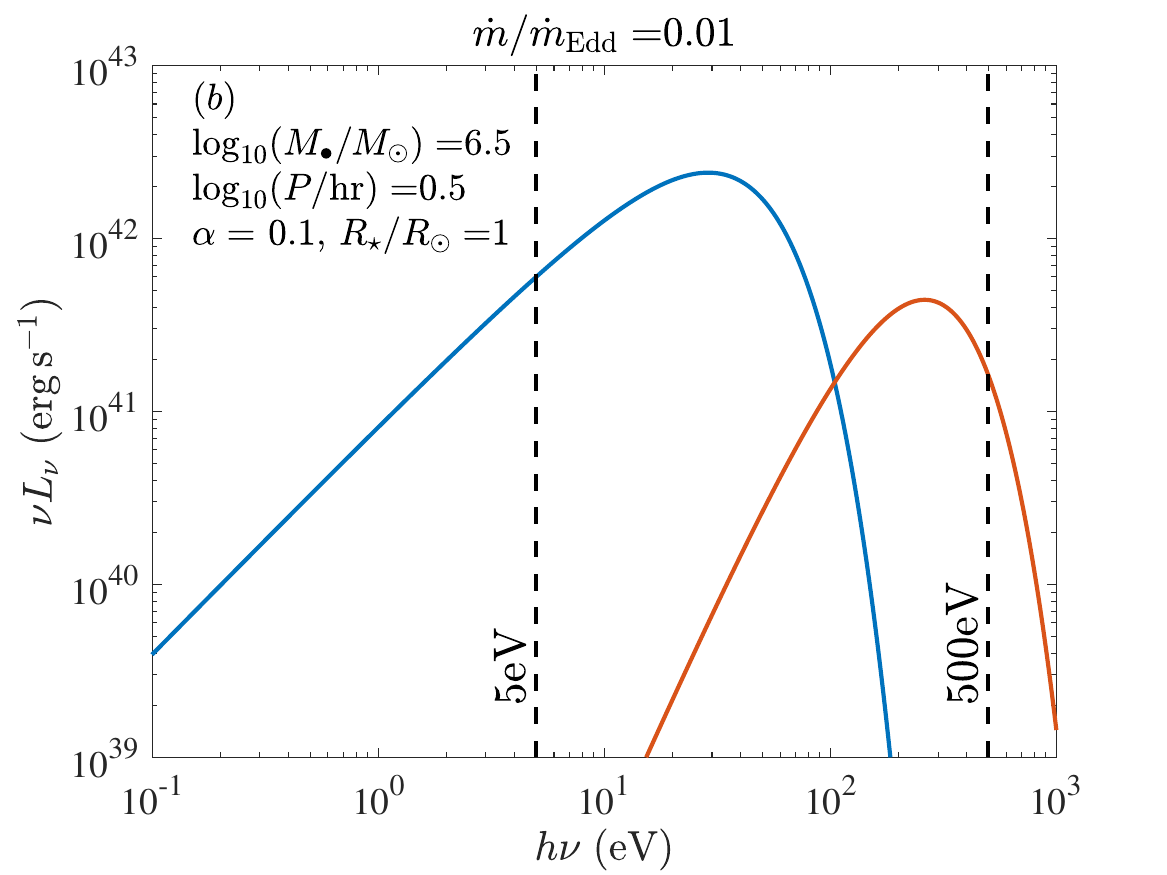}
    \caption{The expected landscape of QPE-like flares, assuming an accretion disk of Eddington ratio $\dot{m}=0.01$ and viscosity $\alpha=0.1$, and inclination $\cos{i}=0.5$. The top-left patch corresponds to flares detectable in soft X-rays, whereas the bottom-right patch is where the flares are visible in UV. Each patch encompasses predictions for a range of stellar radii $R_\star$ in the range $0.5-5 \, \rm R_\odot$. The reference frequencies for UV and X-ray detection were taken to be $h\nu_{\rm obs} = 5\, \rm eV$ and $h\nu_{\rm obs} = \rm 0.5 \, \rm keV$, respectively. The highlighted points, $(a)$ and $(b)$ correspond to the bottom panels, showing the disc and flare SED.  Diagonal dashed lines show contours of constant collision radii $r_0/R_{\rm g}$. }
    \label{fig:detectability}
\end{figure*}

\begin{figure}
    \includegraphics[width=0.49\textwidth]{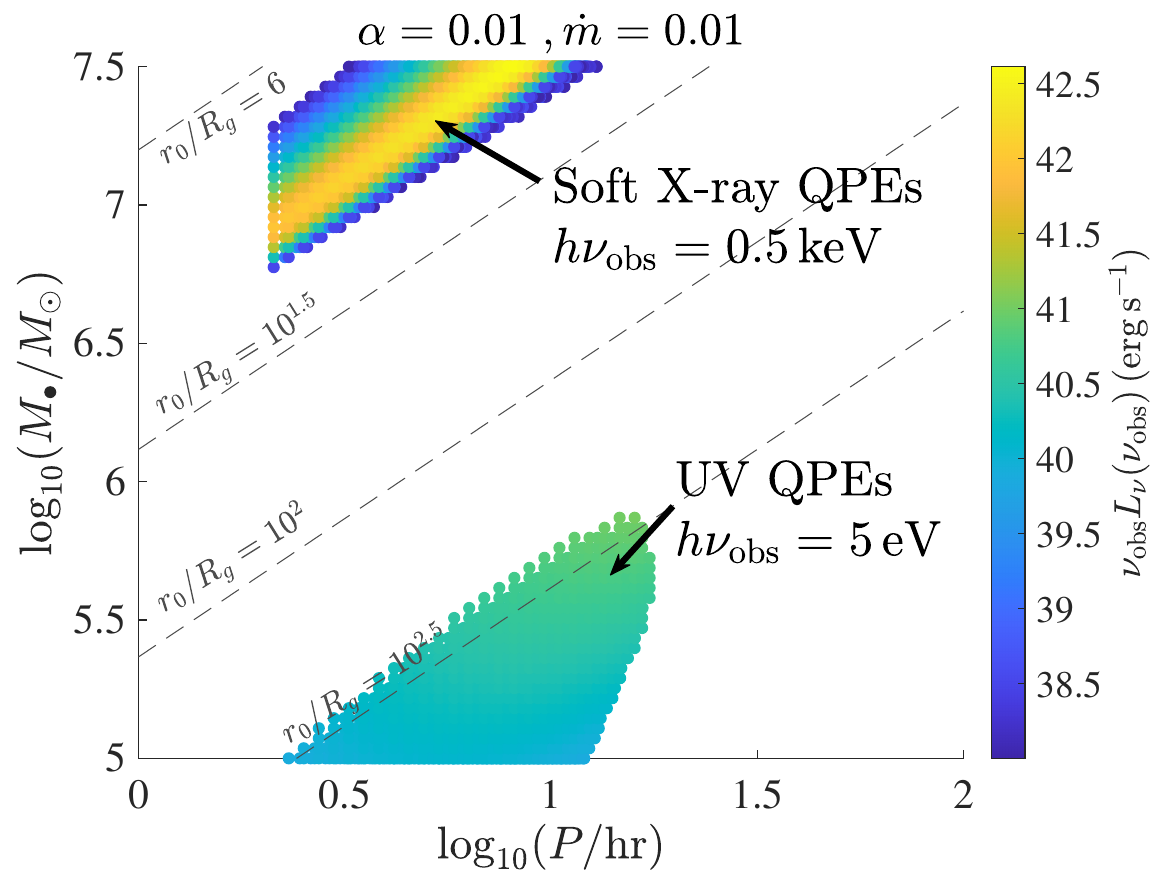}
    \includegraphics[width=0.49\textwidth]{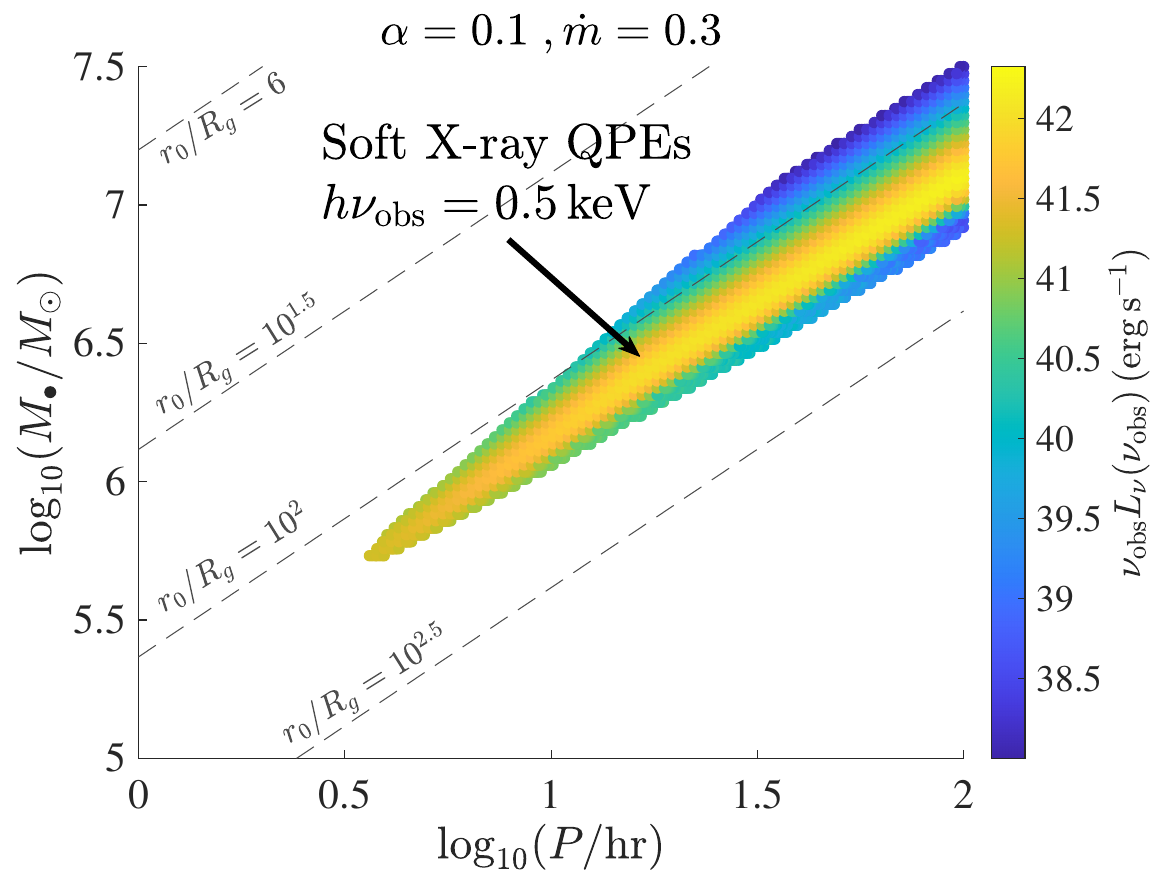}
    \caption{Same as the main panel of Figure \ref{fig:detectability}, but for different assumptions about the accretion rate and viscosity of the disk. The left panel shows the islands of parameter space for which QPEs are detectable in soft X-ray and UV, respectively, for the same Eddington ratio ($\dot{m}=0.01$) as in Fig.~\ref{fig:detectability}, but lower viscosity, $\alpha=0.01$. The right panel corresponds to the same value of $\alpha = 0.1$ as Fig.~\ref{fig:detectability}, but for a higher Eddington ratio ($\dot{m}=0.3$). Here, no UV QPEs are detectable, while X-ray QPEs occur at comparatively lower SMBH masses.}
    \label{fig:detectability_alpha_mdot}
\end{figure}

\begin{figure}
    \centering
    \includegraphics[width=\columnwidth]{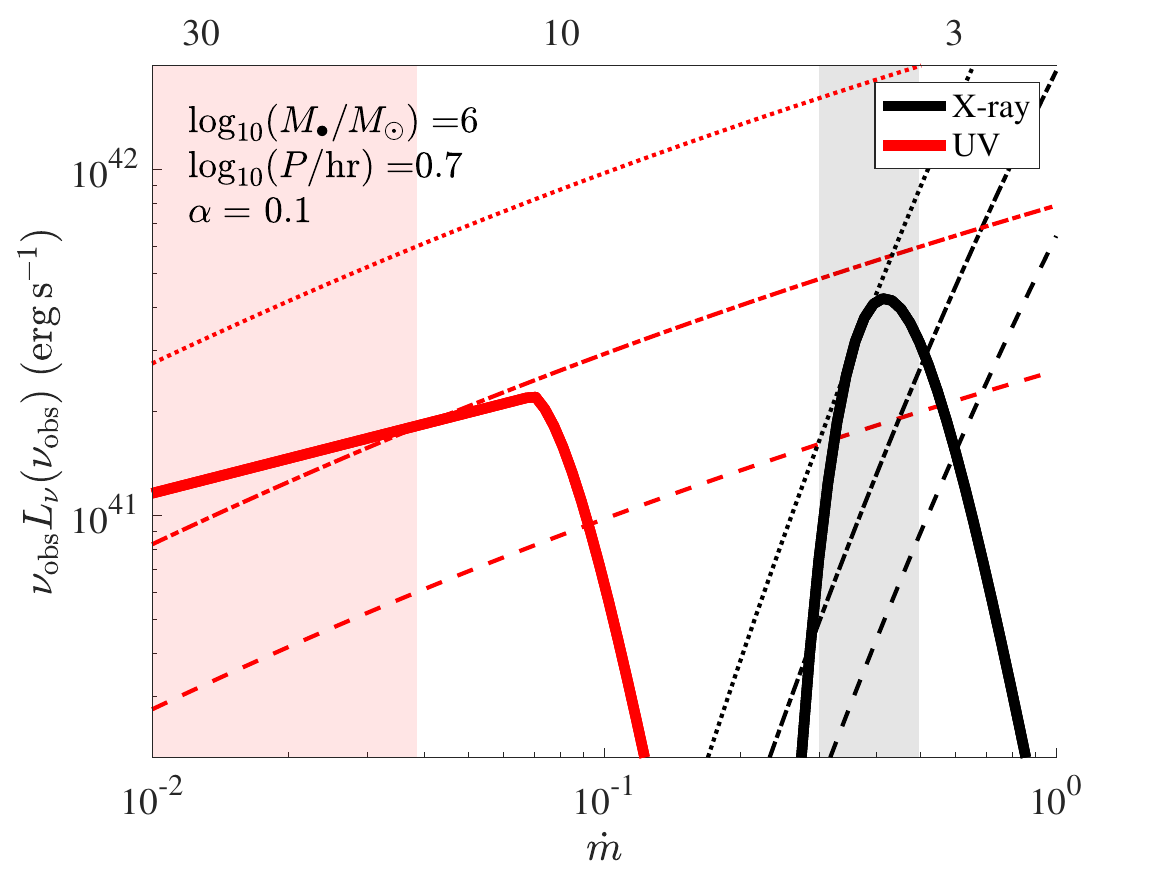}
    \caption{Luminosity of X-ray (\textit{black solid}) and UV (\textit{red solid}) QPE flares as a function of the disc accretion rate $\dot{m}$, for an assumed black hole mass $M_\bullet=10^6 \, \rm M_\odot$, viscosity $\alpha = 0.1$, and fixed stellar orbital period $P_{\rm orb} \approx 2 P_{\rm QPE} \approx 10 \, \rm hr$.  The disk luminosity at the relevant frequency is shown for comparison in the same colour, with \textit{dashed}, \textit{dash-dotted} and \textit{dotted} curves corresponding to different disc inclination angles with respect to the observer's line-of-sight, $\cos{i}= \{0.1,0.3,1\}$, respectively. The top horizontal axis shows the time, in years, since the disruption of a sun-like star, assuming that the disc accretion rate is set by fall-back from a TDE, $\dot{m} \propto t^{-5/3}$. This plot exemplifies how a single orbiting EMRI embedded in a disk with time-evolving accretion rate, could produce at different times, detectable X-ray QPEs, UV QPEs, or potentially both. Specifically, the highlighted gray and pink regions correspond to phases of detectability in X-rays and UV, respectively, assuming $\cos{i}=0.3$. }
    \label{fig:Fixed_MBH_P_vary_mdot}
\end{figure}

\section{Detection Prospects}
\label{sec:detection}

Figure \ref{fig:detectability} summarizes the landscape of observable X-ray and UV QPEs in the parameter space of SMBH mass and QPE period, for an assumed disk accretion rate $\dot{m} = 0.01$, viscosity $\alpha = 0.1$, and range of stellar radii $R_{\star} = 0.5-5 \, R_{\odot}$, corresponding to main-sequence or moderately-evolved stars.  Colored regions of this diagram correspond to those defined as being ``detectable'' according to a combined criteria on the flare temperature (for different assumed UV and X-ray observing frequencies) as well as the requirement that the flare emission outshine that of the quiescent disk (see below). 

 The flare can outshine the disk in one of two ways.  For relatively high SMBH masses and short orbital periods (small $r_0/R_{\rm g} \sim 30$), gas and radiation in the collision ejecta are out of equilibrium ($\eta > 1$; Eq.~\eqref{eq:eta}).  In this regime, the flare emission peaks in the X-rays and hence can stick out above the Wien tail of the softer X-ray emission emitted by the innermost radii of the quiescent accretion disk, as shown in the right bottom panel of Fig.~\ref{fig:detectability}.  This corresponds to the regime of the observed X-ray QPEs (\citetalias{Linial&Metzger23}).  Note that in the right subpanel of Fig.~\ref{fig:detectability} we have plotted the flare SED as a blackbody spectrum of temperature $T_{\rm obs}$, despite the fact that in the photon-starved regime ($\eta>1$), the spectrum can be more complex (e.g., optically-thin free-free spectrum at low frequencies, Comptonized into a Wien spectrum around $h\nu \approx kT_{\rm obs}$; e.g., \citealt{Nakar&Sari10}).  
 
On the other hand, for lower-mass SMBH and longer orbital periods, the radiation and gas are in equilibrium ($\eta < 1$), resulting in flare emission that peaks at lower frequencies $h \nu_{\rm obs} \approx 3k_{\rm B}T_{\rm obs} \approx 5$ eV  in the UV (defined in this paper as the center of the ULTRASAT window), as shown in the bottom left inset panel. Indeed, from Eqs.~\eqref{eq:LQnu},\eqref{eq:Lpk}, we see that
\be
\frac{L_{\rm QPE}}{\nu L_{\rm Q,\nu}} \approx \frac{0.24}{\cos i}\,\mathcal{R}_{\star}^{2/3}M_{\bullet,6}^{-1/3}\dot{m}_{-1}^{-1/3}\mathcal{P}_{\rm QPE,4}^{-2/3}\left(\frac{h\nu}{5\,{\rm eV}}\right)^{-4/3},
\label{eq:flare2diskratio}
\ee
which demonstrates how it is possible for the flare to outshine the UV disk emission for large stellar radii, low SMBH masses, low accretion rates, and/or large viewer inclination angles.  Note that the latter may evolve in time if the disk undergoes precession about the spin axis of the SMBH.  

Even if UV-QPE flares can outshine their disks, to be discovered they must still be luminous enough to be detected in UV survey missions.  The colored contours in Fig.~\ref{fig:detectability} show that while X-ray QPEs are necessarily luminous $L_{\rm QPE} \sim 10^{42}$ erg s$^{-1}$ (consistent with observed X-ray QPE sources), the UV QPE are typically less so ($L_{\rm QPE} \lesssim 10^{41.5}$ erg s$^{-1}$). This follows from the larger collision radii, and hence lower star-disk collision velocities, of the UV-temperature flares.   

To better quantify a detectable luminosity metric, we consider a hypothetical UV survey covering a fraction $f_{\Omega} < 1$ of the sky to a flux depth $F_{\nu, \rm lim}$.  A flare of luminosity $\nu L_{\nu}$ can be observed to a distance
\begin{multline}
    D_{\rm lim} = \left(\frac{L_{\nu}}{4\pi F_{\nu, \rm lim}}\right)^{1/2} \approx 130\,{\rm Mpc}\,\left(\frac{\nu L_{\nu}}{10^{41}\,{\rm erg\,s^{-1}}}\right)^{1/2} \times \\
    \left(\frac{F_{\nu, \rm lim}}{4\times 10^{-29}\rm erg \,s^{-1}\, Hz^{-1} \, cm^{-2}}\right)^{-1/2}.
\end{multline}
Assuming the presence of $f_{\rm QPE}$ QPE sources per Galactic nucleus, the number of detectable sources by the survey is given by
\begin{align}     
    N_{\rm QPE} & \simeq \frac{4\pi}{3}D_{\rm lim}^{3} \, f_{\Omega} \, f_{\rm QPE} \, \mathcal{D}_{\rm eff} \, n_{\rm MW} \nonumber \\
    & \approx 22 \; \mathcal{D}_{\rm eff}\left(\frac{f_{\Omega}}{0.1}\right)\left(\frac{f_{\rm QPE}}{10^{-4}}\right)\left(\frac{\nu L_{\nu}}{10^{42}\,{\rm erg\,s^{-1}}}\right)^{3/2}\times \nonumber \\
    & \left(\frac{F_{\nu, \rm lim}}{4\times 10^{-29} \, \rm erg \,s^{-1}\, Hz^{-1} \, cm^{-2}}\right)^{-3/2},
    \label{eq:NQPE}
\end{align}
where we use the local density of Milky Way-like galaxies $n_{\rm MW} \sim 6\times 10^{6}$ Gpc$^{-3}$ as a proxy for potential QPE hosts.  

Here, we have normalized $f_{\rm QPE} \sim 10^{-5}-10^{-4}$ to a value similar to the range inferred for X-ray QPEs (e.g., \citealt{Metzger+22,Arcodia+22}; R.~Arcodia, private communication), though we note large uncertainties in both this number (being based on a small event sample) and its assumed extension to UV QPEs. While X-ray QPE appear to favor lower-mass galaxies, the occupation fraction of the very low-mass SMBH $M_{\bullet} \lesssim 10^{5.5}M_{\odot}$ we find necessary to generate UV QPE emission, remains uncertain (e.g., \citealt{Greene+20}).  
The inclusion of the QPE duty cycle $\mathcal{D}_{\rm eff} \le 1$ in Eq.~\ref{eq:NQPE} is also conservative for a multi-epoch survey, because given $\gtrsim 1/\mathcal{D}$ observations, the chance of one of the epochs overlapping the flare duration should be of order unity (i.e. $\mathcal{D}_{\rm eff} \simeq 1$). 

ULTRASAT \citep{Sagiv+14} will be sensitive at $\lambda = 230-290$ nm ($h \nu \simeq 4.3-5.4$ eV), reaching a $5\sigma$ sensitivity of 22.4 AB magnitude ($F_{\nu,\rm lim} \approx 4\times 10^{-29}$ erg s$^{-1}$ cm$^{-2}$) across an instantaneous field of view of $\approx 200$ deg$^{2}$ for a 900 second (15 minute) integration.  Most of the observing time will be spent on a high cadence survey, cycling through $\Delta \Omega \sim 6800$ deg$^{2}$ ($f_{\Omega} = 0.16$ of the whole sky) covering 10 fields of view per day (4-day cadence per field).  
ULTRASAT will also complete an all-sky survey that will reach a limiting magnitude of 23 to 23.5 mag ($F_{\nu,\rm lim} \approx 1-2\times 10^{-29}$ erg s$^{-1}$ cm$^{-2}$), but with fewer epochs.
Thus, to detect $N_{\rm QPE} \gtrsim \rm few$ sources with ULTRASAT, from Eq.~\eqref{eq:NQPE} we see that for fiducial assumptions, the flare luminosity must reach $\nu L_{\nu} \gtrsim 10^{41}(10^{42})$ erg s$^{-1}$ for $f_{\rm QPE} \sim 10^{-4}(10^{-5})$, similar to our predictions for UV-QPE sources (Fig.~\ref{fig:detectability}).

The proposed UVEX mission will possess an instantaneous field of view of $\approx 12$ deg$^{2}$, with sensitivity extending also into the FUV ($\lambda = 140-190$ nm; $h \nu = 6.5-8.9$ eV) and reaching an AB magnitude depth of 24.5 for a 900 s integration \citep{Kulkarni+21}, roughly a factor of 6 deeper than ULTRASAT.  Depending on the designed cadence and sky-coverage of its surveys, UVEX would appear somewhat more promising to discover UV-QPE sources than ULTRASAT.

By comparison, the time-domain survey on the now decommissioned GALEX mission had a cadence of 2 days and a baseline of operations of 3 years, but covered only $\sim 40$ deg$^{2}$ (e.g., \citealt{Gezari+13}), leading to $N_{\rm QPE} \ll 1$ for GALEX.  
In principle, star-disk collisions from even larger radii in the disk may give rise to flares peaking at even lower frequencies in the optical range $h\nu_{\rm obs} \lesssim $ 1 eV detectable by optical time-domain surveys such as {\it Zwicky Transient Facility} or the {\it Rubin Observatory}.  However, such optical-QPE flares would be even less luminous, making their detection in optical time-domain surveys challenging.

\section{Discussion and Conclusions}
\label{sec:discussion}

We have demonstrated that models in which X-ray QPEs are produced by disk-star collisions (\citetalias{Linial&Metzger23}), naturally predict that a similar phenomena should extend to flares with lower effective temperatures, which produce emission peaking in the UV instead of the X-ray band.  For otherwise similar stellar radii and disk properties, such ``UV-QPEs'' occur for somewhat larger disk-collision radii, corresponding to longer-period orbits (for similar $M_{\bullet}$) or lower $M_{\bullet}$ (for similar orbital periods).  Although the accretion disk itself is brightest at X-ray energies, X-ray QPE flares are still visible because of the harder radiation they produce as a result of photon starvation.  By contrast, for the conditions which characterize the production UV QPEs, the gas and radiation achieve equilibrium, resulting in softer flare radiation, which are thus somewhat more challenged to outshine the disk. 

The estimates provided in this paper are approximate, and many details and uncertainties enter at the factor of a few level, which could affect whether UV-QPEs are detectable over the quiescent disk in a given system.  These include the detailed shape of the flare light curve; the disk-observer inclination angle and UV spectrum of the quiescent disk (e.g., \citealt{LuQuataert23}); the orientation of the stellar orbit relative to the disk (e.g., star-disk inclination angle; prograde vs.~retrograde) which sets the collision velocity and effective interaction cross-section.  

Nevertheless, we find UV-QPE flare luminosities that at least compete with the disk emission across a wide parameter space (Fig.~\ref{fig:detectability},\ref{fig:detectability_alpha_mdot}).  The flare luminosities achieved in these ``detectable'' regions of parameter space may furthermore be sufficient for the detection of a few such sources to be discovered by impending UV time-domain missions ULTRASAT and UVEX, depending on uncertainties such as the occupation fraction of stellar EMRIs around low-mass SMBH.  In large part a result of the lower collision speeds which give rise to UV-QPEs, we note that disk-star interactions which generate UV-QPEs tend to be less destructive to the star than those responsible for X-ray QPEs (\citetalias{Linial&Metzger23}), possibly enabling a longer lifetime for UV-QPE sources (and hence a higher observed source rate $N_{\rm QPE}$ for a given stellar EMRI formation rate; Eq.~\ref{eq:NQPE}).

In principle, a given stellar EMRI could produce either or both X-ray-QPEs and UV-QPEs, at different times in its evolution.  For example, in a tidal disruption event (as was proposed to precede the QPE activity in GSN-069 and XMMSL1 J024916.6-041244; \citealt{Miniutti+19,Chakraborty+21}), the mass-accretion rate $\dot{m}$ is expected to drop as a function of time, potentially producing X-ray QPEs at early times when $\dot{m} \sim \mathcal{O}(0.1)$ and UV QPEs at later times when $\dot{m}\sim \mathcal{O}(0.01)$ (see Fig.~\ref{fig:Fixed_MBH_P_vary_mdot} for an illustration).  This leads to the testable predictions that systems currently observed as X-ray QPEs may eventually cease to flare in X-rays on timescales of years, but then would eventually \textit{reappear} as UV-QPEs with similar periodicity as the disk accretion rate continues to drop. Within this interpretation, the episode of detectable UV QPEs is expected to span $\sim$decades, roughly $\sim 10$ times longer than the phase of detectable X-rays of the same system (Fig.~\ref{fig:Fixed_MBH_P_vary_mdot}). The difference in durations of the active QPE phases naturally translates to a higher occupation fraction in Galactic nuclei $f_{\rm QPE}$ than the value estimated based on the observed X-ray QPE population.

Likewise, a stellar EMRI that experiences significant radial migration (e.g., due to gravitational wave-radiation or gas-drag due to the disk collisions themselves) between gas accretion events, could produce UV-QPEs at large $P_{\rm QPE}$ and X-ray QPEs upon moving to smaller $P_{\rm QPE}$.  We note that yet shorter orbital periods are unlikely to be sampled by observed X-ray QPEs insofar that the stellar destruction time due to mass ablation from disk collisions is generally shorter than the radial migration time (\citetalias{Linial&Metzger23}).

The appearance, disappearance and recent reappearance of X-ray QPEs in conjunction with secular changes in the quiescent flux has been observed and characterized for the QPE system GSN-069 \citep{Miniutti+19,Miniutti+23,Miniutti+23b}. Our model naturally accounts for the observed trends discussed in \cite{Miniutti+23b}: When the accretion's Eddington ratio exceeds $\dot{m} \gtrsim 0.5$, the inner regions of the disk are hot and bright, outshining the relatively dim flares. At intermediate accretion rates ($0.1 \lesssim \dot{m} \lesssim 0.5$) the flares are visible in X-rays, with an observed temperature exceeding the inner regions of the disk. 
At even lower accretion rates (probably yet to be reached in GSN-069), thermal equilibrium is achieved ($\eta \lesssim 1$) and the flare temperature drops rapidly drops (as $T_{\rm obs} \propto \dot{m}^{5/2})$. Though collisions may no longer be visible in X-rays soon after this transition, at sufficiently low accretion rates $\dot{m} \approx 0.01$, UV flares may outshine the disk emission and become observable.

The QPE candidate XMMSL1J024916.6041244 \citep{Chakraborty+21} has similarly demonstrated long-term evolution of its quiescent X-ray emission, with corresponding changes in the QPE activity. Observations in 2006 reveal soft X-ray QPE activity, which were not detected in followup observations performed in 2021, 15 years later. During these 15 years the source's quiescent flux has decreased by a factor of $\sim 100$, with the disappearance of X-ray QPEs in agreement with our model. Detection of UV variability of the source will confirm the survival of the EMRI that produced the X-ray flares in 2006. Lack of variability may be attributed to flares that are too dim compared with the disc, or the destruction of the EMRI due to drag-driven ablation and/or tidal stripping.

\begin{acknowledgments}
    IL acknowledges support from a Rothschild Fellowship and The Gruber Foundation. BDM was supported in part by the National Science Foundation (grant No. AST-2009255). The Flatiron Institute is supported by the Simons Foundation.
\end{acknowledgments}

\bibliography{UV_QPEs_apjl}
\bibliographystyle{aasjournal}

\end{document}